\newif\ifreview
\newif\ifarxiv
\newif\ifpnas
\newif\ifnature
\newif\ifLineNumbers
\newif\ifxrallowed
\newif\ifbibtexallowed

\arxivtrue

\ifarxiv
    \xrallowedfalse
    \bibtexallowedtrue
    \reviewfalse
    \pnasfalse
    \naturefalse
    \LineNumbersfalse
\fi

\documentclass[
notitlepage,
reprint,
showkeys,
aps,
prx,
citeautoscript,
longbibliography,
nobibnotes,
showkeys,
altaffilsymbol
]{revtex4-2}

\usepackage{amsmath,amssymb}
\usepackage{booktabs}
\usepackage[inline]{enumitem}
\usepackage[pdftex,final]{graphicx}
\usepackage{xcolor}
\usepackage{hyperref}
\usepackage{longtable}
\newcommand{\captionof}[2]{%
  \refstepcounter{#1}%
  \par\noindent\textbf{\csname #1name\endcsname~\csname the#1\endcsname:} #2\par\medskip
}

\graphicspath{{figures/}}
\def\fig{Fig.}
\def\figs{Figs.}
\def\tab{Table}
\def\tabs{Tables}
\def\supp{Appendix}

\begin{document}

\title{Size and shape of terrestrial animals}
\author{Neelima Sharma}
\email{neelima.sharma@ucl.ac.uk}
\affiliation{Department of Cell \& Developmental Biology, University College London, London, UK, WC1E 6BT}
\author{Madhusudhan Venkadesan}
\email{m.venkadesan@yale.edu}
\affiliation{Department of Mechanical Engineering \& Materials Science, Yale University}

\begin{abstract}
Natural selection for terrestrial locomotion has yielded unifying patterns in the body shape of legged animals, often manifesting as scaling laws~\cite{Huxley1931NotesDifferentialGrowtha,Thompson1942fp,McMahon1973tx, Alexander1979aa, Alexander1981, Alexander1985cb, Biewener1989ab,Kaspari1999,Vanhooydonck1999mq, Usherwood:2013dq, Allen2013aa,schmidt1984scaling,biewenerBiomechanicalConsequencesScaling2005,pelabonEvolutionMorphologicalAllometry2014}.
One such pattern appears in the frontal aspect ratio.
Smaller animals like insects typically adopt a landscape frontal aspect ratio, with a wider side-to-side base of support than center of mass height.
Larger animals like elephants, however, are taller than wide with a portrait aspect ratio.
Known explanations for postural scaling are restricted to animal groups with similar anatomical and behavioural motifs~\citep{Kaspari1999,Allen2013aa,Usherwood:2013dq,McMahon1973tx,Alexander1979aa,Biewener1989ab,Alexander1981,Alexander1985cb,Vanhooydonck1999mq}, but the trend in frontal aspect ratio transcends such commonalities.
Here we show that vertebrates and invertebrates with diverse body plans, ranging in mass from 28\,mg to 22000\,kg, exhibit size-dependent scaling of the frontal aspect ratio driven by the need for lateral stability on uneven natural terrain.
Because natural terrain exhibit scale-dependent unevenness~\cite{Goodchild:1982fk, Mandelbrot:1983kx, Brown1985, moore1991, Power1991aa, With1994aa, Shepard:2001kx}, and the frontal aspect ratio is important for lateral stability during locomotion~\citep{Kubow1999,Henry2001, Daley2018cv,Kirby1987,Beloozerova2003,Voloshina2013xk,Donelan2004,Bauby2000,Hak2012,Hof2010,Musienko2014limb,Voloshina2015sk,Druelle2019small,dhawale2023human}, smaller animals need a wider aspect ratio for stability.
This prediction is based on the fractal property of natural terrain unevenness, requires no anatomical or behavioural parameters, and agrees with the measured scaling despite vast anatomical and behavioural differences.
Furthermore, a statistical phylogenetic comparative analysis found that shared ancestry and random trait evolution cannot explain the measured scaling.
Thus, our findings reveal that terrain roughness, acting through natural selection for stability, likely drove the macroevolution of frontal shape in terrestrial animals.
\end{abstract}
\maketitle

\begin{figure}[!hbt]
    \centering
    \includegraphics[width=\columnwidth,keepaspectratio]{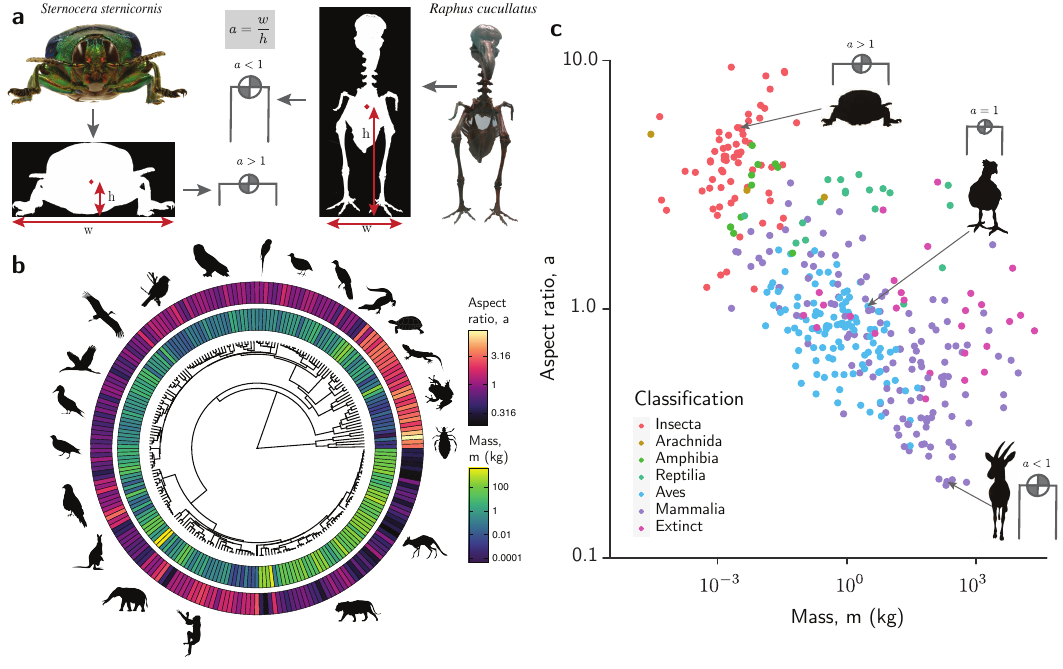}
    \caption[Aspect ratio decreases with mass]{
        {\bfseries Aspect ratio and mass of terrestrial animals.}
        {\bf a,} Aspect ratio $a$ of the species is the ratio of the base of support $w$ to the height $h$ of the center of mass above ground.
        {\bf b,} Phylogeny, aspect ratio $a$, and mass $m$ of 364 terrestrial species spanning vertebrates and invertebrates.
        {\bf c,} Mass versus aspect ratio, on a log-log scale, showing that animals get narrower as they get heavier.
    }
\label{fig:figure1}
\end{figure}

\section{Main text}\label{main-text}

Scaling of body and limb geometries with size lends a window into how physical factors have shaped the body through natural selection \citep{schmidt1984scaling,biewenerBiomechanicalConsequencesScaling2005,pelabonEvolutionMorphologicalAllometry2014,Thompson1942fp,Huxley1931NotesDifferentialGrowtha}.
Specifically, the scaling of body posture with size is thought to reflect selection for terrestrial locomotion in several animal groups.
Ants exhibit non-isometric scaling of leg length with body size, which is thought to arise from habitat driven selection for finding footholds on foliage covered ground~\citep{Kaspari1999}.
Birds and related archosaurs evolved to shift their center of mass cranially and also decreased in size, possibly due to simultaneous terrestrial and aerial locomotion demands \citep{Allen2013aa}.
Locomotion energetics is thought to affect whether vertebrates with dynamically similar gaits deviate from geometric similarity in their leg posture \citep{Usherwood:2013dq}, but additional explanations include bone stress limits \citep{McMahon1973tx,Alexander1979aa}, muscle stress management \citep{Biewener1989ab}, and departure from geometric similarity in muscle mass distribution and tendon moment-arms \citep{Alexander1981}.
These explanations invoke commonalities in morphological and behavioural motifs such as shared habitats, limb anatomy, or gait patterns, and therefore do not translate to the scaling of frontal shape that transcends such commonalities~(figure~\ref{fig:figure1}).

We estimated the frontal aspect ratio $a$ of 364 species, spanning invertebrates and vertebrates ranging in mass from 28\,mg to 22000\,kg (figure~\ref{fig:figure1}a,b, table~\ref{tab:extndtab1}), and found that it decreases with mass $m$ with a statistically significant linear relationship between $\log_{10} m$ and $\log_{10} a$~(figure~\ref{fig:figure1}c, mean slope $\pm$ standard error $=-0.13\pm{}0.01$, R$^2=0.41$, $p<0.0001$, \supp~\ref{sec:phylogenyanalysis}).
Smaller, lighter animals are more sprawled and landscape shaped when seen head-on ($a>1$), but larger, heavier animals are more slender and portrait shaped ($a<1$).
The linear regression suggests a power-law, but the scatter in the data precludes precisely estimating the exponent and could obscure non-power-law relationships.
Nevertheless, the data show a clear relationship between size (mass) and frontal aspect ratio despite vast differences among these animals in habitat use, behavioural traits, anatomy, and developmental trajectories.
Because frontal aspect ratio profoundly affects lateral stability~\citep{Kubow1999,Henry2001, Daley2018cv,Kirby1987,Beloozerova2003,Voloshina2013xk,Donelan2004,Bauby2000,Hak2012,Hof2010,Musienko2014limb,Voloshina2015sk,Druelle2019small,dhawale2023human}, we investigated whether stability could be the underlying driver of the measured scaling.

\begin{figure}[!ptbh]
    \centering
    \includegraphics[width=0.906\columnwidth,keepaspectratio]{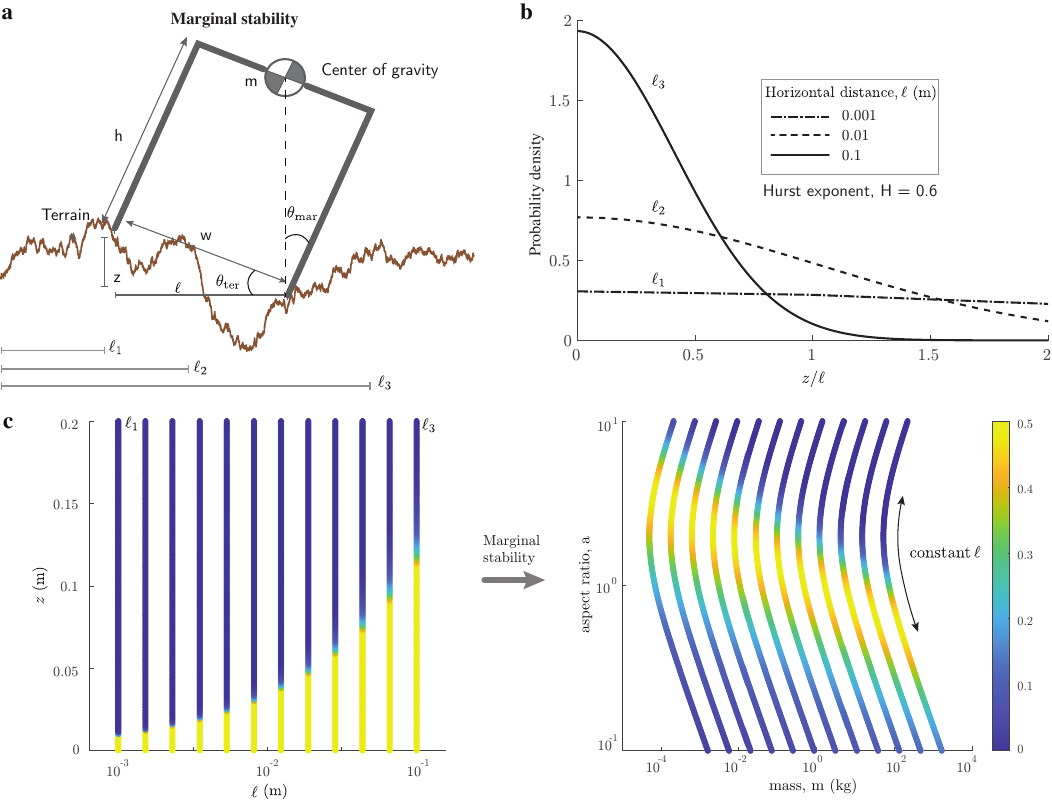}
    \caption[Effect of width on stability]{{\bfseries Effect of width on stability.}
    {\bf a,} Definitions of base of support width $w$, center of mass height $h$ for an animal of mass $m$ that is laterally tilted by $\theta_{\rm ter}=\tan^{-1}(z/\ell)$ when the ground support points are horizontally separated by $\ell$ and experience a height difference \(z\).
    The point of marginal stability is when $\theta_{\rm ter}=\theta_{\rm fall}$, i.e.\ the center of mass resides exactly above one of the support points.
    {\bf b,} The distribution $z/\ell=\tan\theta_{\rm ter}$, for typical natural terrain with Hurst exponent $H=0.6$ shows that natural terrain appear rougher at smaller lengths (wider distribution).
    {\bf c,} Mapping of $p_{Z}(z;\ell)$ univariate distributions for ten values of $\ell$ to the corresponding constant $\ell$ curves in the $(\log a, \log m)$ space.
    The calculation is performed with $H=0.6$ and $D=0.0135$.
    }%
    \label{fig:figure2}
\end{figure}

The frontal aspect ratio dictates the maximum lateral tilt beyond which an animal would tip over laterally and lose stability.
For an animal of width $w$ and height $h$, the aspect ratio is $a=w/h$ and the angle at the margin of falling over is given by $\theta_{\rm fall} = \tan^{-1}(a/2)$ (figure~\ref{fig:figure2}a).
Despite the complications of how stability is achieved in dynamic locomotion~\citep{Alexander2002StabilityManoeuvrabilityTerrestriala,Jindrich2009ManeuversLeggedLocomotiona,Aoi2016AdvantageStraightWalka,Haagensen2022ExploringLimitsTurninga}, the passive notion of lateral stability applies broadly; wider implies improved lateral stability not only when standing still \citep{Kirby1987,Henry2001,Beloozerova2003}, but also when walking \citep{Voloshina2013xk,Donelan2004,Bauby2000,Hak2012,Hof2010,Musienko2014limb} or running \citep{Kubow1999,Voloshina2015sk,Druelle2019small,Daley2018cv,dhawale2023human}.
However, being wider, and thus more stable, comes at multiple costs~\citep{Alexander2002StabilityManoeuvrabilityTerrestriala}.
Greater lateral width increases muscle and tissue stresses and also metabolic energy consumption \citep{Usherwood:2013dq,Biewener1989ab,Voloshina2013xk,Donelan2004}.
Further, it may also compromise manoeuvrability because body tilt is an important means to execute turns~\citep{Alexander2002StabilityManoeuvrabilityTerrestriala,Haagensen2022ExploringLimitsTurninga,Aoi2016AdvantageStraightWalka,Jindrich2009ManeuversLeggedLocomotiona}.
The static frontal profile sets a baseline around which neural control of the limbs can further modulate the width, and thus the frontal aspect ratio.
Having a marginally stable passive shape, at the cusp of tipping over, is a trade-off that allows active neural control of the limbs to achieve a wider, stable stance or a narrower, more manoeuvrable one.
Therefore, we ask how the frontal aspect ratio of the passive body shape should scale with size to maintain marginal stability on uneven terrain.

On natural terrain, it is the ground's unevenness that tilts the body.
Therefore, understanding the relationship between body width and ground-induced tilt is part of the stability analysis.
The ground-induced tilt is $\theta_{\rm ter} = \tan^{-1}(z/\ell)$, where $z$ is the vertical height difference between two ground support points separated horizontally by a distance $\ell$~(figure~\ref{fig:figure2}a).
Stability, so that the animal does not tip over due to the ground-induced tilt, implies $\theta_{\rm fall} \ge \theta_{\rm ter}$, i.e.\ $a/2 \ge z/\ell$.
Marginal stability is when the two angles are equal.
But natural terrain exhibit a random height profile.
Specifically, the random height variations of natural terrain are captured by a fractional Brownian process model such that the standard deviation $\sigma_z$ of the random height difference $z$ scales as $\sigma_z \propto \ell^{H}$, with a Hurst exponent $H<1$~(\supp~\ref{sec:fbm terrain})\cite{Goodchild:1982fk, Mandelbrot:1983kx, Brown1985, moore1991, Power1991aa, With1994aa, Shepard:2001kx}.
Thus, the marginally stable aspect ratio $a$ scales with size $\ell$ as,
\begin{equation}
    a \propto \ell^{(H-1)}.\label{eqn:aspect ratio scaling with ell}
\end{equation}
Typical values of $H$ range from 0.3 to 0.8 and is always less than 1, implying that smaller the size $\ell$, larger the $z/\ell$, and so natural terrain appear rougher when zoomed-in~(figure~\ref{fig:figure2}b, extended table~\ref{tab:extndtab2})~\cite{Goodchild:1982fk, Mandelbrot:1983kx, Brown1985, moore1991, Power1991aa, With1994aa, Shepard:2001kx}.
Thus, the exponent $(H-1)$ in equation~\eqref{eqn:aspect ratio scaling with ell} is always negative for natural terrain, resulting in the prediction that the smaller an animal is the more landscape-oriented it should be.

Testing the predicted scaling law requires estimates of animal size and its aspect ratio across a diverse set of species.
The aspect ratio is dimensionless and can be estimated from head-on images without requiring size calibration.
But estimates of size in terms of mass $m$ are more readily available than head-on, length-calibrated images.
Therefore, to gather a large dataset for testing the predicted scaling law, we developed an approach that does not rely on length-calibrated images by deriving a mapping between the animal's width $w$ and height $h$ to its mass $m$ and aspect ratio $a$ (Methods~\ref{sec:methods for probability of being marginally stable on natural terrain}).
The mapping from $(w,h)$ to $(m,a)$ is possible because mass is proportional to the product $w\times h$, whereas the aspect ratio is equal to $w/h$, thus yielding two independent equations.

Thus, for a given horizontal separation $\ell$, the probability of encountering a height variation $z$ maps to the probability of the animal parameterized by the pair $(m,a)$ being marginally stable, i.e.\ with $a=2z/\ell$.
Thus, the hypothesis of marginal stability transforms the terrain's one-parameter family of probability density functions $p_{\rm ter}(z;\ell)$ into the one-parameter family of probability densities $p_{\rm fall}(\{m,a\};\ell)$ for an animal of mass $m$ and aspect ratio $a$ to be marginally stable (figure~\ref{fig:figure2}c, Methods, \supp~\ref{sec:computing marginal animal density function}).

\begin{figure}[!bht]
    \centering
    \includegraphics[width=0.55\columnwidth,keepaspectratio]{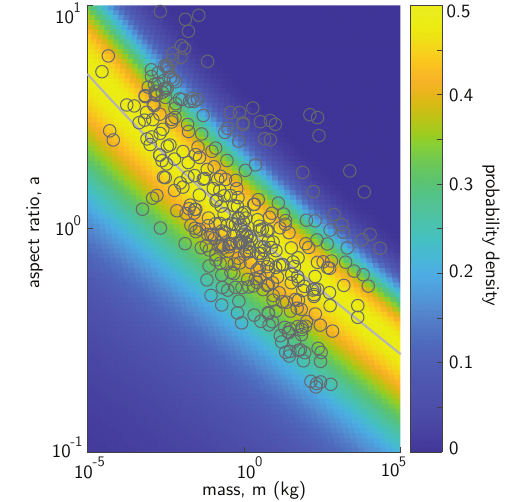}
    \caption[Probability density of aspect ratio and mass.]{{\bfseries Probability density of aspect ratio $a$ and mass $m$} (heatmap) overlaid with data (grey circles). The grey curve is the loci of the mode of the density at each mass $m$, representing the most likely scaling law.
    }
    \label{fig:figure3}
\end{figure}

Our measurements of 364 animals, ranging in mass from 28~mg to 22000~kg, are overlaid on the predicted probability density (figure~\ref{fig:figure3}).
The prediction is based on the statistical unevenness properties of natural terrain alone, with only one fitting parameter that translates the overall curve but does not affect its slope (\supp~\ref{subsec:params}).
The slope arises from the scaling of aspect ratio with body size (equation~\eqref{eqn:aspect ratio scaling with ell}), which is governed solely by the Hurst exponent $H$ of terrain unevenness.
A general allometric exponent converts from size to mass, but introduces no anatomical or behavioural assumptions specific to the animals in the dataset.
Furthermore, the probabilistic scaling law allows for a comparison with the mean trend, as is possible with typical deterministic scaling laws \citep{schmidt1984scaling,biewenerBiomechanicalConsequencesScaling2005,pelabonEvolutionMorphologicalAllometry2014}, as well as the variability across animals that is to be expected in natural settings.
Most measurements are clustered around the mode of the predicted density function (grey curve and its yellow tinted neighbourhood in figure~\ref{fig:figure3}), and with departures from the mode occurring more often towards wider, more stable animals than the other way round.
This is consistent with the premise that passive stability that is attuned to the statistics of natural terrain is important for terrestrial animals.

Caution is warranted in interpreting the correspondence between the predicted and measured allometry as evidence for adaptive evolution.
Trait correlations might arise from phylogenetic relatedness rather than shared selective pressures because closely related species are not independent data points~\cite{Felsenstein1985aa,Grafen1989aa}.
We used Phylogenetic Generalized Least Squares (PGLS) to test whether the relationship persists after accounting for shared ancestry~(Methods~\ref{phylogenetic-generalized-least-squares-analysis}, \supp~\ref{sec:phylogenyanalysis})\citep{pelabonEvolutionMorphologicalAllometry2014,Symonds2014,Pagel1994,Pagel1999,Grafen1989aa}.
The PGLS regression used a Brownian motion model with a $\lambda$ transformation to account for phylogenetic covariance in the residuals.
This regression estimated a slope of $-0.08$ between $\log a$ and $\log m$ and remains significant ($p=2\times{}10^{-8}$, \tab~\ref{tab:pglslambda}).
This is shallower than the ordinary least squares estimate of $-0.13$.
But if the scaling were predominantly an artifact of shared ancestry, the PGLS slope would approach zero.
Instead, it remains significantly different from zero, showing that allometry persists over and above the effects of shared ancestry.

To further assess whether the correlation could arise from independent trait evolution, we simulated 100,000 instances of Brownian evolution on the phylogenetic tree (Methods).
When mass and aspect ratio evolve independently (no cross-correlation), the probability of observing a slope as steep as $-0.08$ is only $3 \times 10^{-5}$.
By contrast, when the traits co-evolve with cross-correlations matching those estimated from our data, this probability is $0.63$.
Thus, shared ancestry and random drift do not explain the observed scaling.
Rather, the two traits have evolved interdependently, consistent with selection for lateral stability on uneven terrain.

The terrain-based scaling law captures the distribution of aspect ratio across 364 species spanning eight orders of magnitude in mass, establishing lateral stability as a critical factor in frontal shape evolution.
Adaptation to specialized niches or multiple environments, however, could involve trade-offs that underlie some of the observed variability.
For example, semi-aquatic and semi-aerial species require a streamlined body to present a slender cross-section to minimize fluid drag, branches and leaves in arboreal habitats are generally not orthogonal to gravity, and it remains unknown whether burrows are characterized by a fractional Brownian process when considering fossorial species.
Furthermore, specialized adaptations such as adhesive pads in insects, claws in reptiles, and prehensile tails in mammals can stabilize terrestrial locomotion through means other than the base of support.
We exclude species that obligatorily inhabit aquatic, arboreal, or subterranean regions, as well as species that rely primarily on adhesive pads for stability rather than base of support, but include those species that traverse loose and sandy substrates where adhesion is not effective.
As a result of these habitat variations and anatomical specializations, the Hurst exponent of terrain is not universal and the allometric relationship between body dimensions and mass is approximate.
Nevertheless, the negative scaling of aspect ratio with mass holds across the full range of terrestrial Hurst exponents from $H = 0.3$ to $0.8$ (complete sensitivity analyses in \fig~\ref{fig:extndfig3}a, \figs~\ref{fig:extndfig5}b and~\ref{fig:extndfig5}c, \S\ref{sec:variability}).
Despite all these exceptions and sources of variability, the measurements cluster around the mode predicted from lateral stability considerations and exhibit the scaling predicted by the need for stability on natural terrain.

Selection for lateral stability that is attuned to the unevenness of natural terrain underlies the macroevolution of frontal body shape across eight orders of magnitude in mass from arthropods to megafauna.
This terrain-based driver of body shape adds to the current understanding of environmental factors that have shaped the evolutionary diversification of terrestrial animals~\citep{Shubin2009jb, McKenna2021zo}.
An ant navigating pebbles and an elephant among boulders face similar stability challenges because Earth's fractal terrain has shaped both their aspect ratios.

\section{Methods}
\label{sec:methods}

\subsection{Data collection}
\label{subsec:data-collection}

Aspect ratios were collected using published photographs or by
taking frontal profile photographs of museum specimens.
We relied on scale free frontal photographs of the animals to measure their aspect ratios because data on the absolute dimensions of animals is severely limited and at best consists only of measurements of body length~\cite{moura2023phylogeny}.
Photographs were either collected using Nikon D810 Digital Camera with Nikon 52 mm lens (Nikon Corp., Japan) or Samsung Galaxy S9 SM-G960U1 with a 12MP OIS camera (Samsung Electronics America).
We processed the images using Adobe Photoshop (21.0.1 Release, Adobe, San Jose, CA) to delete image background.
Fiji Version 1.0 \citep{Schindelin2012} was used to convert the animal photograph with no background to an 8-bit grayscale image, and further converted into black and white by adjusting the thresholds so that the animal was represented with a value of 255 and the background with a value 0.
We measured the centroid height $h$ from the black and white photograph, and the width of the base of support $(w)$ by finding the horizontal distance from the tip of the left end of the left leg to the tip of the right end of the right leg using a bounding box.
We assume that the animal has uniform density, and therefore, the centroid and center of mass overlap.
Using $w$ and $h$, we measure the aspect ratio of the animal as $a=w/h$.
See \fig~\ref{fig:figure2}c for a pictorial depiction of image analysis.

We used published values of animal mass for a subset of species (Extended Data Table~\ref{tab:extndtab1}).
For some arthropods that were collected from the field, we measured mass using Mettler AT261 DeltaRange Analytical Balance (Mettler-Toledo, LLC, Columbus, OH).
In the case of dry arthropod specimens obtained from private collection, we measured the insect dry weight and multiplied it by $10/3$ to convert it into live weight, based on the studies that show that the water content in insects is $\sim 70$\% of the live weight \citep{Studier1992}.

\subsection{Probability of being marginally stable on natural terrain}%
\label{sec:methods for probability of being marginally stable on natural terrain}

The conditions for an animal of width $w$ and center of mass height $h$ to be marginally stable when it stands on uneven terrain with the ground contact points horizontally separated by distance $\ell$ and with a height difference $z$ are,
\begin{subequations}
    \begin{align}
        \ell^2 + z^2 &= w^2\ \text{for ground contact and} \label{eqn:animal Pythogoras}\\
        \frac{w}{2 h} &= \frac{z}{\ell}\ \text{for marginal stability}.
    \end{align}%
    \label{eqn:standing and marginal stability}
\end{subequations}
Solving these equations would yield animal shape parameters $w$ and $h$ as functions of $z$ and $\ell$.
We transform the parameters that determine animal shape to more easily measurable aspect ratio and mass by invoking the definitions that the aspect ratio $a$ is the ratio of width $w$ to height $h$ and mass is proportional to the volume $w\times h\times d$.
Furthermore, the fore-aft dimension $d$ can be eliminated because it scales as the cube root of mass, $d = ({m}/{\rho_0})^{1/\beta}$ \citep{Niklas:1994eu} (where $\rho_{0}$ is an empirically derived constant).
Thus, we obtain,
\begin{subequations}
    \begin{align}
        a &= \frac{w}{h}\ \text{and}\\
        m &= \rho d w h = \rho \left(\frac{m}{\rho_0}\right)^{1/\beta} w h.
    \end{align}%
    \label{eqn:aspect ratio and mass}
\end{subequations}
Solving~\eqref{eqn:standing and marginal stability} and~\eqref{eqn:aspect ratio and mass}, for $a$ and $m$ in terms of $z$ and $\ell$ we find,
\begin{subequations}
    \begin{align}
        a &= \frac{2 z}{\ell},\ \text{and}\\
        m &= {\left(\frac{\rho}{2 \rho_0^{1/\beta}}\right)}^{{\beta}/{(\beta-1)}} {\left(\frac{\ell}{z}(\ell^2+z^2)\right)}^{{\beta}/{(\beta-1)}}.
    \end{align}
\end{subequations}
For natural terrain, $z$ is a random variable~\citep{Goodchild:1982fk, moore1991, Mandelbrot:1983kx, Molz1997, With1994aa}.
The probability that the height difference between two points separated by a horizontal distance $\ell$ lies in the interval $[z, z+dz]$ is given by $p_{\rm ter}(z;\ell)\,dz$ where,
\begin{equation}
    p_{\rm ter}(z;\ell) = \sqrt{\frac{2}{\pi D \ell^{2H}}}\,\exp\left(-\frac{z^2}{2 D \ell^{2H}}\right),\ z \ge 0.%
    \label{eqn:PDF of terrain}
\end{equation}
Equations~\eqref{eqn:standing and marginal stability} and~\eqref{eqn:aspect ratio and mass} transform the random variable $z$ and thus
map the one-parameter family of probability densities $p_{\rm ter}(z;\ell)$ associated with the terrain to another family of probability densities $p_{\rm fall}(\{\log m, \log a\};\ell)$ associated with the marginally stable animal (figures~\ref{fig:figure2}e,~\ref{fig:figure3}).
The detailed procedure for computing the transformed probability density function $p_{\rm fall}(\{\log m, \log a\}; \ell)$ and visualizing it are in \supp~\ref{sec:computing marginal animal density function}.

\paragraph*{Parameter values.} We use independently estimated parameter values for all but $D$, which was found by fitting the most likely scaling to our data.
Based on the exponent of the allometric relationship between fore-aft length and mass found by Niklas \citep{Niklas:1994eu}, we picked $\beta = 3$ and the pre-factor to set $\rho_0 = 0.77$ kg \citep{Niklas:1994eu}.
Terrestrial animals are generally denser than water, and we picked $\rho = 1250$ kg m$^{-3}$ following the choice made in previous modelling work for aquatic animals minus their swim bladders \citep{alexanderSizeSpeedBuoyancy1990}.
Finally, for the terrain Hurst exponent, we chose $H=0.6$ as the central estimate for terrain over length scales that are appropriate to terrestrial animals (\tab~\ref{tab:params}).
The value of $D = 0.0135$ m$^{2(1-H)}$ is obtained by minimizing the residuals, which are defined as the square of the distance between the mode and the data point along a constant $\ell$ curve (\supp~\ref{subsec:params}).

\subsection{Scaling law based on muscle stress considerations}
\label{scaling-law-based-on-muscle-stress-considerations}

Muscles support the bodyweight of the animal.
Whereas body weight increases as the volume of the animal, muscle tension increases only as muscle's cross-sectional area, which is proportional to the square of the animal's linear dimension.
For maintaining similar peak musculoskeletal stresses in the frontal profile, the muscles must support the animal's body weight.
Thus, the moment due to ground reaction force about the limb joints must be supported by the moment due to the muscles.
Thus, we have $F_{m}.r = F_{g}.\frac{w}{2}$ where $r$ is the
moment arm of the muscle about the limb joint, $w/2$ is the moment arm
of the ground reaction force about the joints, $F_m$ is the muscle
tension and $F_g$ is the force due to gravity.
The muscle's moment arm scales as the bone's diameter, and from elastic buckling considerations \citep{McMahon1973tx}, bone diameter $d$ and bone length $l$ are related by $l \sim d^{2/3}$.
We assume that the length of the bone scales as the height of the animal's center of mass $h$, which scales as $h\propto m^{1/3}$ \citep{Niklas:1994eu, Pontzer2007}.
Muscle tensions scale as the cross-sectional area $A_m$ of the muscle,
related to body weight as $A_m\propto m^{0.8}$ \citep{Alexander1981}.
Using these relationships in the equation, that moment due to gravity must equal the moment applied by the muscles across the joints in the front profile, the upper bound between the aspect ratio $a=w/h$ and animal mass $m$ is $a \propto m^{-1/30}$.

\subsection{Ordinary and Phylogenetic Regression Analysis}
\label{phylogenetic-generalized-least-squares-analysis}

We constructed a phylogenetic tree for 222 species with known branch lengths out of 364 species using Timetree \citep{Kumar2017}.
We used \texttt{drop.tip} from \texttt{ape} function to construct a pruned phylogenetic tree that considered extant and unique species to generate a rooted ultrametric tree.
We tested the effect of animal size $\log m$ on their aspect ratio $\log a$ using ordinary linear regression.
Diagnostics on the residuals showed that the assumptions for linear regression were not met (Table~\ref{tab:linearregression} and Fig.~\ref{fig:extndfig4}a).
We performed a phylogenetic generalized least squares (PGLS) analysis to include the effect of phylogenetic relatedness.
First, we tested the phylogenetic signal in individual traits according to different models of evolution including Brownian motion (BM), Ornstein-Uhlenbeck (OU), and Early Burst (EB) models.
A Brownian Model (BM) of evolution with $\lambda$-transformation best described the evolution of individual traits, as depicted by the lowest Akaike Information Criterion (AIC) scores (Table~\ref{tab:individualsignal}).
Therefore, we performed Phylogenetic Generalized Least Squares (PGLS) regression between \(\log a\) and \(\log m\) with Brownian motion models of evolution using maximum likelihood estimate (MLE) of $\lambda$, $\kappa$, and $\delta$ transformations (Table~\ref{tab:bmcomparison}).
The $\lambda$ transformation compresses the internal branches of the phylogenetic tree,
the $\delta$ transformation models the variation in evolution rates by raising the elements of the covariance matrix to $\delta$,
and in $\kappa$ transformation, all the branch lengths are raised to the power $\kappa$.
An AIC score difference of 4 or more is typically considered for preferring one model over another, and therefore, $\lambda$-transformation was chosen for parsimoniousness (Table~\ref{tab:bmcomparison}).
We used the package \texttt{caper} in RStudio to perform the PGLS analysis.

We use Monte Carlo simulations of multivariate trait co-evolution on the phylogenetic tree for the species used in our study to estimate the slope between aspect ratio and mass.
This analysis is an alternative to PGLS and relaxes any assumption on the distribution of the residuals.
Two models were compared; the first model considers the presence of evolutionary cross-correlations and the second model excludes them.
We estimated the evolutionary variance-covariance matrix using
\texttt{ratematrix} \citep{Caetano2017} that uses a Bayesian Markov
chain Monte Carlo approach to estimate the evolutionary correlation
among traits.
We find the variances, $\sigma^2(\log a) = 0.0002$ and $\sigma^2(\log m) = 0.0052$, and the cross-covariance, $\textrm{cov}(\log a, \log m) = -0.0004$.
The phylogenetic tree was rescaled by Pagel's $\lambda$ transformation, as described above, to apply the branch length transformation similar to the PGLS analysis.
Using \texttt{fastBM} from \texttt{geiger} library in RStudio, we generated a null hypothesis for the evolution of $\log a$ and $\log m$ as two independent traits using 100,000 Monte Carlo simulations.
Using the function \texttt{simulate-bm-model} in the \texttt{castor} library, we implemented multivariate trait co-evolution of $\log a$ and $\log m$ where cross-correlations were considered.
The initial (ancestral state) at the root is assumed to be $\log a = 0$ and $\log m = 0$, depicting an animal with a square profile and mass of 1 kg.
We compared the probability of obtaining the slope predicted by the PGLS regression from the two analyses (Fig.~\ref{fig:extndfig4}b).
Detailed results are present in \fig~\ref{fig:extndfig4}, and \tabs~\ref{tab:linearregression}, \ref{tab:individualsignal}, \ref{tab:bmcomparison}, \ref{tab:pglslambda}.

\subsection{Software}

MATLAB 9.12.0.1884302 (R2022a) was used to calculate the probability densities and RStudio version 4.1.3 was used to perform the phylogenetic analysis.

\subsection{Acknowledgements}

We thank M. Dickman for sharing their private collection, L. Gall and K. Zyskowski for help with collections at the Yale Peabody Museum, and J. Sartore, P. Lanoue, S. Kessel, and K. Rudloff for sharing photographs. Additional images were obtained from the Field Museum of Natural History and the Macaulay Library. M.V. was supported by the National Science Foundation (Award No. 2046120). N.S. was supported by Yale's Integrated Graduate Program in Physical and Engineering Biology, the Pierre W. Hoge Foundation Fund, and the Alpheus B. Stickney Scholarship Fund. 

\bibliographystyle{apsrev4-2}
\bibliography{references}

\clearpage
\onecolumngrid
\appendix

\section{Probability densities of the shape and size of marginally stable animals}
\label{sec:computing marginal animal density function}

\subsection{Model of the animal}
\label{model-of-the-animal}

An animal of mass $m$ is modelled to have center of mass height
$h$, base of support $w$, and fore-aft length $d$.
With average density $\rho$, empirically estimated exponent $\beta$,  and constant $\rho_0$ based on previously found allometry across animals \citep{Niklas:1994eu}, the aspect ratio $a$ and mass $m$ are related to the dimensions $w$ and $h$ by,
\begin{subequations}
\begin{eqnarray}
a &=& \frac{w}{h} \label{eqn:aspect ratio def supp}, \\
m &=& \rho d h w \label{eqn:mass of the animal supp},\ \text{and}\\
d &=& \left(\frac{m}{\rho_0}\right)^{1/\beta}. \label{eqn:niklas empirical eqn supp}
\end{eqnarray}%
\label{eqns:animal governing equations supp}
\end{subequations}

\subsection{Model of the terrain}
\label{model-of-the-terrain}

The terrain is modelled as a one-dimensional fractional Brownian (fBm) process $b(x)$ where $x$ is the horizontal coordinate~(\S\ref{sec:fbm terrain}).
The increments, defined as the difference in the fBm process $b(x)$ between two points horizontally spaced by distance $\ell$, are modelled as fractional Gaussian noise $y(x,\ell)$.
The distribution of increments is modelled as a Gaussian process with mean zero and variance that depends on the horizontal distance $\ell$ as $\sigma^2 = D \ell^{2H}$.
The prefactor $D$ is a constant that is equal to the standard deviation of $y$ for a unit increment and $0<H<1$ is the fractional Hurst exponent that describes the roughness of the terrain.
The symmetry of the frontal shape of the animal poses a symmetry in the frontal tilt due to the terrain, such that the height difference $-y$ and $y$ pose the same risk of falling.
Therefore, we consider the absolute value of the height increments, $z=|y|$.
Transformation of the Gaussian random variable $y$ provides a folded-normal distribution for $z$ (Eq.~\ref{eqn: fz chi supp}) with mean, $\mu_z = \sigma \sqrt{2}/\sqrt{\pi}$, and variance, ${\sigma_z}^{2} = \sigma^{2} (1-2/\pi)$,
\begin{equation}
\text{pdf of height increments: } p_{\rm ter}(z;\ell) = \sqrt{\frac{2}{\pi \sigma^2}} \exp\left(-\frac{z^2}{2 \sigma^2}\right), z>0, \sigma = \sqrt{D} \ell^{H}.
\label{eqn: fz chi supp}
\end{equation}
When an animal stands on the uneven terrain, the horizontal distance $\ell$ is the projected base-width of the animal.
Therefore, the absolute increment or the height difference $z$, and the base width $w$, and the distance $\ell$ are related by,
\begin{equation}
\text{Definition of standing: } w^2 = z^2 + \ell^2.
\label{eqn:standing def supp}
\end{equation}

\subsection{Probability densities associated with marginally stable animals}
\label{probability-density-in-the-log-a-log-m-space}

To maintain a baseline lateral stability on earth's terrain, we propose that the tipping angle associated with marginal stability of the animal ${\theta}_{\rm fall}$, defined by its aspect ratio, equals the tilt induced by the terrain $\tan{\theta_{\rm ter}}$ (\fig~\ref{fig:figure2}b),
\begin{equation}
\tan{\theta}_{\rm fall} = \tan{\theta_{\rm ter}}
\implies \frac{a}{2} = \frac{z}{\ell}
\label{eqns:critical ratio supp}
\end{equation}
Using Eq.~\ref{eqns:animal governing equations supp}, \ref{eqn:standing def supp} and \ref{eqns:critical ratio supp}, we find that the aspect ratio and mass for constant values of $\ell$ are related as,
\begin{equation}
\log m = \frac{\beta}{\beta-1} \log \frac{\rho \ell^2 (a^2 + 4)}{4 \rho_0^{1/\beta} a}.
\label{eqn:am constant l relationship supp}
\end{equation}
The mapping of one-parameter family of density functions $p_{\rm ter}(z;\ell)$ (Eq.~\ref{eqn: fz chi supp}) to a one-parameter family of density functions in $p_{\rm fall}(\{\log m, \log a\}; \ell)$ space gives the probability densities of $\log$-transformed aspect ratio and mass of marginally stable animals as $\ell$ is varied  (\fig~\ref{fig:extndfig1} and \fig~\ref{fig:figure2}).
The probability density of animal with the $\log$-transformed aspect ratio and mass for marginal stability is given by,
\begin{equation}
p_{\rm fall}(\{\log a,\log m\};\ell) = \sqrt{\frac{2}{4 \pi D \ell^{2(H-1)}}} \exp\left(-\frac{a^{2}}{8 D \ell^{2(H-1)}}\right) a.
\label{eqns:pdf gamma supp}
\end{equation}

\subsection{Locus of modes of the probability densities associated with marginally stable animals}
\label{locus-of-modes-of-the-distribution}

The locus of the modes of the family of univariate distributions parameterized by $\ell$ provides a deterministic scaling relationship between aspect ratio and size.
We obtain the mode of univariate distributions from the maxima of the probability density function,
\begin{subequations}
\begin{gather}
\frac{d p_{\rm fall}(\{\log m, \log a\};\ell)}{d \log a} = 0\\
\implies \log a = \log(2 \sqrt{D} \ell^{H-1}). \label{eqn:mode1 supp}
\end{gather}
\label{eqns:mode supp}
\end{subequations}
Using the relationship between $m$ and $\ell$ from Eq.~\ref{eqn:am constant l relationship supp} and Eq.~\ref{eqn:mode1 supp}, we obtain the locus of the modes of the distributions in implicit form in terms of the variables $a$ and $m$ as,
\begin{equation}
m = \left(\frac{\rho}{2^{\frac{2H}{H-1}} D^{\frac{1}{H-1}} \rho_0^{\frac{1}{\beta}}}\right)^{\frac{\beta}{\beta-1}} \left(a^{\frac{3-H}{H-1}}(a^2+4) \right)^{\frac{\beta}{\beta-1}}
\label{eqn:locus of modes supp}
\end{equation}
The asymptotes of the predicted scaling when $a\ll 1$, that is a highly portrait profile, and for $a\gg 1$, that is a highly landscape profile, are
\begin{subequations}
\begin{gather}
a\gg 1 \implies a \propto m^{\frac{(H-1)(\beta-1)}{\beta(H+1)}} \\
a\ll 1 \implies a \propto m^{\frac{(H-1)(\beta-1)}{\beta(3-H)}}
\end{gather}
\label{eqns:asymptotes supp}
\end{subequations}

\subsection{Parameter values}
\label{subsec:params}

The parameter values and their sources are provided in \tab~\ref{tab:params}.
There is one fitting parameter $D$ that corresponds to the horizontal shift of the map.
We minimize the sum of residuals between the data points and the scaling law obtained as the locus of the modes to find the value of parameter $D$.
The residuals are defined as the sum of the squared distance of the data points from the mode of the distribution along the constant $\ell$ curves.
Each $(a_i, m_i)$ data pair where $i\in{1,2,\cdots,364}$ for 364 animals defines constant values of $\ell_i$ for each data point pair according to
\begin{equation}
\ell_i = \sqrt{\frac{4 \rho_0^{1/\beta} a_i m_i^{1 - 1/\beta} }{\rho (a_i^2+4)}}
\label{eqn:value of ell supp}
\end{equation}
Along the $\ell_i$ curve, we measure the distance of the data point from the mode of the distribution.
The distance $s_i$ between two points along the constant $\ell_i$ curve is given by
\begin{subequations}
\begin{gather}
s_i = \int\limits_{(\log a)_{\rm mode}(\ell_i)}^{(\log a)_i } \sqrt{\left(\frac{d \log a}{d \log a}\right)^2 + \left(\frac{d \log m}{d \log a}\right)^2} d \log a \\
\text{where } \frac{d \log m}{d \log a} = \frac{\beta}{\beta-1}\frac{a^2 - 4}{a^2 + 4}, \\
\text{and } (\log a)_{\rm mode}(\ell_i) = \log{2 \sqrt{D} {\ell_i}^{H-1}}.
\end{gather}
\label{eqns:value of D supp}
\end{subequations}
The residuals are defined as the square of the distance between the mode and the data point along a constant $\ell$ curve, and so, we have $r_i = s_i^2$.
We find the value of $D$ for which $R = \sum_{i=1}^{364} r_i$ is minimized for various values of $H$ (\tab~\ref{tab:Dvalues}).
We obtain a minimum residual value of $D=0.0135$ for $H=0.6$, and present the results for $H=0.6$ in the main paper and the supplement.

\begin{table}[!ht]
\centering
\caption[Parameter values.]{Parameter values.}
\begin{tabular}{ll*{4}{c}}
\toprule
Parameter & Magnitude & Units & Reference \\
\midrule
$\beta$ & $\approx$ 3 & --   & \cite{Niklas:1994eu} \\
$\rho_0$  & 0.77  & kg m$^{-\beta}$ & \cite{Niklas:1994eu} \\
$\rho$  & 1250  & kg m$^{-3}$ & \cite{alexanderSizeSpeedBuoyancy1990} \\
$H$     & (0,1)     & --               & Extended Data Table~\ref{tab:extndtab2} \\
$D$     & 0.0016 &  m$^{2(1-H)}$  & fit \\
\bottomrule
\end{tabular}
\label{tab:params}
\end{table}

\begin{table}[!ht]
    \centering
    \caption[Fitting D values]{For various values of the Hurst exponent $H$, the fitted values of $D$ obtained by minimizing the residuals.}
    \begin{tabular}{ll*{4}{c}}
    \toprule
    H & D (m$^{2(1-H)}$) & R \\
    \midrule
    0.2  & 0.0008 & 530.82 \\
    0.3  &  0.0016 & 425.26 \\
    0.4  &  0.0034 & 353.06 \\
    0.5  &  0.0068& 310.28 \\
    0.6  &  0.0135 & 292.82 \\
    0.7  &  0.0269 & 297.94 \\
    0.8  &  0.0558 & 326.01 \\
    0.9  &  0.1212 & 379.29 \\
    \bottomrule
    \end{tabular}
    \label{tab:Dvalues}
\end{table}

A change in exponent H captures the variability in the terrains of different roughness and over varying length scales, and explains partial spread in the data (\fig~\ref{fig:extndfig3}a).
Although the exact function form changes, the trend is preserved.
For low H, a steep negative slope of $\log a$ versus $\log m$ for lower masses and a shallower slope at higher masses indicates that the roughness drops sharply with an increase in the length scale at smaller scales compared to a more gradual decline at larger scales, and vice versa for larger H.

\subsection{Verification using Monte Carlo simulations}
\label{subsec:monte carlo}

We propagate the distribution of the terrain height increments through the governing equations to find the distribution of aspect ratio and mass in the $\log$-space.
The terrain height perturbations $z$ are random variables picked from a folded-normal distribution with parameter $\sigma$ which depends on the value of $\ell$.
For each $\ell$, the random variables contribute to a random variable pair $(a,m)$.
The dependence of aspect ratio and mass on the random variable $z$ under the stability hypothesis are provided by the following equations:
\begin{subequations}
\begin{gather}
a =  \frac{2z}{\ell}\\
m = \left(\frac{\rho \ell (z^2 + \ell^2)}{2 \rho_0^{1/\beta}z}\right)^{\beta/(\beta-1)}
\end{gather}
\label{eqns:monte carlo supp}
\end{subequations}
We perform Monte Carlo simulations using a $100,000$ realizations of the random variable $z$ for each $\ell$, and by varying $\ell$ over a $1000$ values.
We plot the $(a,m)$ pairs obtained from propagating $z$ in the $\log$-space.
We remind you that visualization in $\log$-space is associated with $\log$-transformed variables.
For a constant $\ell$ curve, we color code the pairs based on the density of points in the bin defined along the $t$ axis to ensures that the results from the Monte Carlo simulations reflect the probability densities obtained analytically in \S\ref{sec:computing marginal animal density function} and \fig~\ref{fig:extndfig3}b.

\section{Phylogenetic comparative analysis}
\label{sec:phylogenyanalysis}

\begin{table}[!ht]
    \centering
    \caption[Relationship between $\log_{10} a$ and $\log_{10} m$ from ordinary least squares (OLS) regression.]{Relationship between $\log_{10} a$ and $\log_{10} m$ for the 222 species considered in our phylogenetic analysis demonstrated by ordinary least squares (OLS) regression.}
    \begin{tabular}{ll*{4}{c}}
    \toprule
     & Estimate & Standard Error & t-value & p-value \\
    \midrule
    Intercept & -0.02491  &  0.01550  &  -1.608   &   0.109  \\
    Slope & -0.13244    &  0.01069  & -12.385   &  $<$2e-16 \\
    \midrule
    R$^2$ = 0.41 & & & & \\
    F-statistic: 153.4 on 1 and 220 DF\\
    \bottomrule
    \end{tabular}%
    \label{tab:linearregression}
    \end{table}

\begin{table*}[htbp]
    \centering
    \caption[Estimating the phylogenetic signal in individual traits according to the BM, OU, and EB models of evolution to inform the choice of the best-fit model.]{Phylogenetic signal in individual traits according to the Brownian motion (BM), Ornstein-Uhlenbeck (OU), and Early Burst (EB) models of evolution. The estimates of $\lambda$ for the BM model of evolution are based on maximum likelihood. The lowest Akaike Information Criterion (AIC) score or the largest log-likelihood score is used to select the best-fit model.}
    \begin{tabular}{lrrrrrlrrr}
        \toprule
      (Aspect ratio)& {AIC} & {LogLik} &            & (Mass) & {AIC} & {LogLik} \\
      \midrule
      BM   & 33.83 & -14.92 &           & BM     & 617.15 & -306.57 \\
      BM ($\lambda = 0.95$)  & \textbf{-102.78} & 54.39 &           & BM ($\lambda=0.98$)  & \textbf{565.98} & -279.99 \\
      OU   & -12.83 & 9.41 &          & OU   & 604.37 & -299.18 \\
      EB     & 35.84 & -14.92 &           & EB    & 619.15 & -306.57 \\
      \bottomrule
\end{tabular}%
\label{tab:individualsignal}%
\end{table*}%

\begin{table*}[htbp]
    \centering
    \caption[Phylogenetic Generalized Least Squares (PGLS) regressions using Brownian Motion (BM) model of evolution for testing the effect of various branch transformations.]{Phylogenetic Generalized Least Squares (PGLS) regression analysis using Brownian Motion (BM) model of evolution to test the effect of $\lambda$, $(\lambda,\delta)$, and $(\lambda,\delta,\kappa)$ transformations. Because the AIC scores are within 4 units of each other, no model is strongly preferred over another.}
      \begin{tabular}{llrr}
        \toprule
      Model  & Parameter estimates &{LogLik} & {AIC} \\
      \midrule
      ($\lambda$) & $\lambda$ = 0.897 &68.57 & -133.15 \\
      ($\lambda$, $\delta$) & $\lambda=0.931$, $\delta=3$ &69.80 & -135.60 \\
      ($\lambda$, $\delta$, $\kappa$) & $\lambda=0.933$, $\delta=3$, $\kappa=1.128$  & 70.09 & -136.19 \\
      \bottomrule
      \end{tabular}%
    \label{tab:bmcomparison}%
\end{table*}

\begin{table}[h]
    \centering
    \caption[Relationship between $\log{a}$ and $\log{m}$ using PGLS regression with $\lambda$-transformation.]{Relationship between $\log{a}$ and $\log{m}$ using PGLS regression with $\lambda$-transformation. Maximum likelihood estimation yields $\lambda$ = 0.897 ($95.0\% {~\textrm CI} : (0.791, 0.952)$).}
      \begin{tabular}{lrrrr}
        \toprule
            & {Estimate} & {SE} & {tvalue} & {pvalue} \\
            \midrule
      (Intercept) & 0.30 & 0.22 & 1.36 & 0.17 \\
      $\log m$ & -0.08 & 0.01 & -5.82 & 2e-08 \\
      \bottomrule
      \end{tabular}%
    \label{tab:pglslambda}%
\end{table}

\section{Sources of variability}
\label{sec:variability}

In our predictions of the scaling law and the probability distributions, the sources of variability are either due to measurement errors, partial data, or imprecise values of parameters in our analysis. We elaborate on these sources below:
\begin{enumerate}
    \item {\it Measurement errors}\/: There are two sources of measurement errors. Although every attempt was made to select frontal profile photographs, the animal photographs used for measurement may not be perfectly head-on, leading to slight errors in the measured aspect ratio. Secondly, animal mass and aspect ratio are not constants but have a range for a given species, but we use a single measurement for the aspect ratio from the photograph and either measure or use published value of mass from the literature. We did not attempt to collect descriptive statistics of animal mass and aspect ratios for each species because a higher precision in these data, which would have varied less than an order of magnitude, would not have contributed to a greater precision in the proposed scaling, which relies on interspecific comparison over eight orders of magnitude. In other words, we expect the variation in aspect ratio and mass of a species to be considerably smaller than the relevant scales in the problem.
    \item {\it Partial data}\/: Our data are limited by the available resources in museums, private collections, available photographs, and local animal species. As a result, the collected data is not comprehensive. Despite this limitation, with 364 species, we capture the scaling in aspect ratio with animal size over eight orders of magnitude and relate it to the terrain unevenness under the stability hypothesis.
    \item {\it Errors associated with the parameter values}\/:
    \begin{enumerate}
\item Presence of adhesive pads at the distal end of the limbs in animals such as insects would provide them with additional stability (\fig~\ref{fig:extndfig5}a). However, adhesive pads are not useful on terrains comprised of loose material such as soil and sand.
\item The analysis of Niklas (1994) \citep{Niklas:1994eu} relies on modeling animals (n=67) as cylinders, and measures the scaling of mass with the longitudinal dimension of the cylinder. For arthropods and quadrupedal vertebrates, the longitudinal dimension is equal to the depth (the third dimension) considered in our model but not for bipedal vertebrates, where fully extended head height is the longitudinal dimension. Thus, the value of the parameters $\rho_0$ and $\beta$ obtained from \citep{Niklas:1994eu} are valid only for invertebrates and quadrupedal vertebrates but not for bipeds. We present the data sorted by the animals for whom the parameter values $\rho_0$ and $\beta$ are valid based on the methods followed by \citep{Niklas:1994eu} (\fig~\ref{fig:extndfig5}b). In our analysis, we use the same value for $\rho_0$ for bipedal vertebrates. It is likely that a different value of $\rho_0$ is obtained when considering the scaling of mass with the depth $d$ of the animal. However, a sensitivity analysis of our scaling to $\beta$  and $\rho_0$ shows that the trend is preserved (\fig~\ref{fig:extndfig5}c).
\item The value of Hurst exponent measured from different terrains and over different increments over the surface of the earth shows that $H$ varies between 0 and 1, as expected from the definition of the fractional Brownian motion. We show that changing the value of $H$ does not impact the predicted scaling between aspect ratio and mass such that smaller animals need to be sprawling and larger animals can be upright to maintain stability on natural terrains (\fig~\ref{fig:extndfig3}a).
\item The value of constant $D$ found by finding the best fit between the scaling and the data agrees with the measured values of $D$ on the uneven terrains. However, the value of $D$ varies over the surface of the earth \citep{Butler2001,Gallant1994} and will introduce some variability in the scaling prediction.
    \end{enumerate}
\end{enumerate}

\section{Upper bound based on muscle stress considerations}
\label{sec:upper bound}

Relying on the empirical observation that peak skeletal stresses remain uniform in mammals despite the decreased scaling of muscle cross sectional area with body weight, Biewener~\citep{Biewener1989ab} argues that similar peak bone and muscle stresses are achieved by a change in locomotor limb posture from crouched to upright with an increase in size.
In the lateral view, with an upright posture, animal limbs are more closely aligned with the ground reaction forces in comparison to a crouched posture, thus reducing the muscle stresses.
Here, we use Biewener's~\citep{Biewener1989ab} arguments to find an upper bound on the aspect ratio of the animal by extending the reasoning to the frontal body plan (\fig~\ref{fig:extndfig6}a).
We argue that in order to maintain similar peak muscle and bone stresses, the frontal aspect ratio of the animal must decrease with size such that small animals can sprawl and crouch, but larger animals need to be upright.

Muscles supports the body weight of the animal.
Therefore, from \fig~\ref{fig:extndfig6}a, we have $F_{m}.r = F_{g}.\frac{w}{2}$ where $r$ is the moment arm of the muscle about the limb joint, $w/2$ is the moment arm of the ground reaction force about the joints, $F_m$ is muscle tension, and $F_g$ is the force due to gravity.
The moment arm of the muscle scales as the radius (or diameter) of the bone, and from elastic buckling considerations~\citep{McMahon1973tx}, bone diameter $d$ and bone length $l$ are related by $l \sim d^{2/3}$.
We assume that the length of the bone scales as the height $h$ of the animal, that in turn scales with body weight as $h\propto m^{1/3}$~\citep{Niklas:1994eu,Pontzer2007}.
Muscle tensions scales as the cross sectional area $A_m$ of the muscle, that is related to body weight as $A_m\propto m^{0.8}$~\citep{Alexander1981}.
Using these relationships, the upper bound between the aspect ratio $a=w/h$ and animal mass $m$ is $a \propto m^{-1/30}$ (\fig~\ref{fig:extndfig6}b).

\section{Tutorial on modelling terrain as a fractional Brownian process}
\label{sec:fbm terrain}

We present a brief tutorial on the modeling of the terrains. The material and its organization is heavily inspired from Molz et al.~\cite{Molz1997} and must be referred for more details on terrain modeling.

\paragraph{Scale invariance.} A function $f(x)$ is scale invariant when the function $f(cx)=c^{\Delta}f(x)$ for some scale factor $c$ and for some exponent $\Delta$. The shape of the function $f(cx)$ is simply a rescaled version of $f(x)$. Scaling here implies linear or affine scaling. In other words, a linear or affine magnification or contraction in $c$ will reshape the function $f(cx)$ to become indistinguishable from $f(x)$. Scale invariance leads to self similarity when $\Delta=1$ and self affinity when $\Delta\ne 1$, because the scale-invariant object looks the same upon such rescaling.

\paragraph{Statistical self similarity.} A real valued stochastic process $b(x)=\{B(x),x\in\mathbb{R}\}$ is self-similar if the distribution of $b(cx)$ for some $c>0$ is identical to the distribution of the the process $c^{H}b(x)$. The exponent $H$ defines the index of self-similarity. The statement of statistical self-similarity in mathematical terms is given by:
\begin{equation}
\{B(cx),x\in \mathbb{R}\} \triangleq \{c^{H}B(x),x\in \mathbb{R}\}
\label{eqns:statistical ss}
\end{equation}

\subsection{Modeling of fractional Brownian processes}

The one dimensional form of fractional Brownian process $b(x)$ (fBm) is defined as a single-valued continuous function, which is defined using stationary, Gaussian, and self-similar increments $y(x)$, also known as fractional Gaussian noise (fGn). The properties of fGn are discussed below.

\paragraph{Stationary increments.}
Consider two points horizontally spaced by an $\ell$ along the fBm: $b(x)$ and $b(x+\ell)$. We define the increment as the difference of the fBm process between these horizontally spaced points, $y(x,\ell) = b(x+\ell) - b(x)$. Stationarity of increments implies that the moments of $y(x,\ell)$ are independent of $x$ and that the probability distribution of $y(x,\ell)$ depends only on $\ell$. Therefore, $\forall x$ and fixed $\ell$, we have mean and variance of the increments as
\begin{subequations}
\begin{gather}
E[y(x,\ell)] = c_1(\ell)\\
E[y(x,\ell)^2] = c_2(\ell)
\end{gather}
\label{eqns:stationary increments}
\end{subequations}

\paragraph{Gaussian increments.}
The Gaussian increments are defined to have the following mean and variance.
\begin{subequations}
\begin{gather}
E[y(x,\ell)] = c_1(\ell) = 0\\
E[y(x,1)^2] = c_2(1) = D \label{eqn:def of D supp}
\end{gather}
\label{eqns:gaussian increments}
\end{subequations}

\paragraph{Self similar increments.}
The increments are self similar with index $H$. For fixed $n$ and $\ell$,
\begin{subequations}
\begin{gather}
y(x, n\ell) \triangleq n^{H} y(x,\ell), n \ge 0, 0<H<1 \\
\implies E[y(x,n\ell)] = E[n^{H} y(x,\ell)] = 0 \label{eqn:mean of increments supp}\\
\text{and } E[y(x,n\ell)^2] = E[n^{2H} y(x,\ell)^2] = n^{2H} E[y(x,\ell)^2] \label{eqn:variogram supp}
\end{gather}
\label{eqns:self similar increments}
\end{subequations}

From equations \eqref{eqn:def of D supp} and \eqref{eqn:variogram supp}, we find the variance of the increments as
\begin{equation}
E[y(x,\ell)^2] = \ell^{2H} E[y(x,1)^2] = D\ell^{2H}
\label{eqns:eqn of variance supp}
\end{equation}

\subsection{fGn increments lead to fBm process}

Here, we present the proof that fractional Gaussian noise increments lead to a fractional Brownian process.
Expanding equations \eqref{eqn:mean of increments supp} and \eqref{eqn:variogram supp} in terms of the fractional Brownian process by using the definition of the increments, we have
\begin{subequations}
\begin{gather}
E[b(x+n\ell) - b(x)] = n^{H} E[b(x+\ell) - b(x)] \\
\text{and } E[(b(x+n\ell)-b(x))^2] = n^{2H} E[(b(x+\ell)-b(x))^2]
\end{gather}
\label{eqns:proof 1}
\end{subequations}
Because the increments are stationary, without loss of generality, we choose $x=0$ where $b \equiv 0$ and obtain,
\begin{subequations}
\begin{gather}
E[b(n\ell)] = n^{H} E[b(\ell)] \\
\text{and } E[b(n\ell)^2] = n^{2H} E[b(\ell)^2]
\end{gather}
\label{eqns:proof 2}
\end{subequations}
which is the definition of a statistically self-similar process.

\subsection{Hurst exponent and terrain roughness}

For $H$ approaching 0, the variance in height increments at shorter distance is larger in comparison to that at longer distances (\fig~\ref{fig:extndfig7}).
In contrast, a surface with Hurst exponent approaching 1 has a high correlation between nearby points.
For natural terrains, $H$ typically varies from 0.3 to 0.8 (Extended Data Table~\ref{tab:extndtab2}).
Our study shows that the lateral stability on terrain requires small animals to have a higher aspect ratio whereas allows large animals to be more slender.
Here, we analyze how a change in Hurst exponent affects the relationship between aspect ratio and mass by plotting the locus of models of the distributions according to equation~\eqref{eqn:locus of modes supp}.
We observe that the change of the exponent conserves the trend such that smaller animals sprawl and larger animals could be slender with a varying dependence on animal mass (\fig~\ref{fig:extndfig3}a).
However, the slope becomes shallower as the value of $H$ increases (\fig~\ref{fig:extndfig3}a).
If stability were the only criteria and if $H=1$, then the width of the base of support would become independent of animal mass as the terrain would approach equal roughness at all lengthscales.
Our results suggest that the presence of niche terrains associated with different Hurst exponents may underlie some of the observed variability and spread in the aspect ratios because niche environments may pose unique selection pressures.

\section*{Extended Data Figures}

\begin{center}
  \includegraphics[width=0.6\textwidth,keepaspectratio]{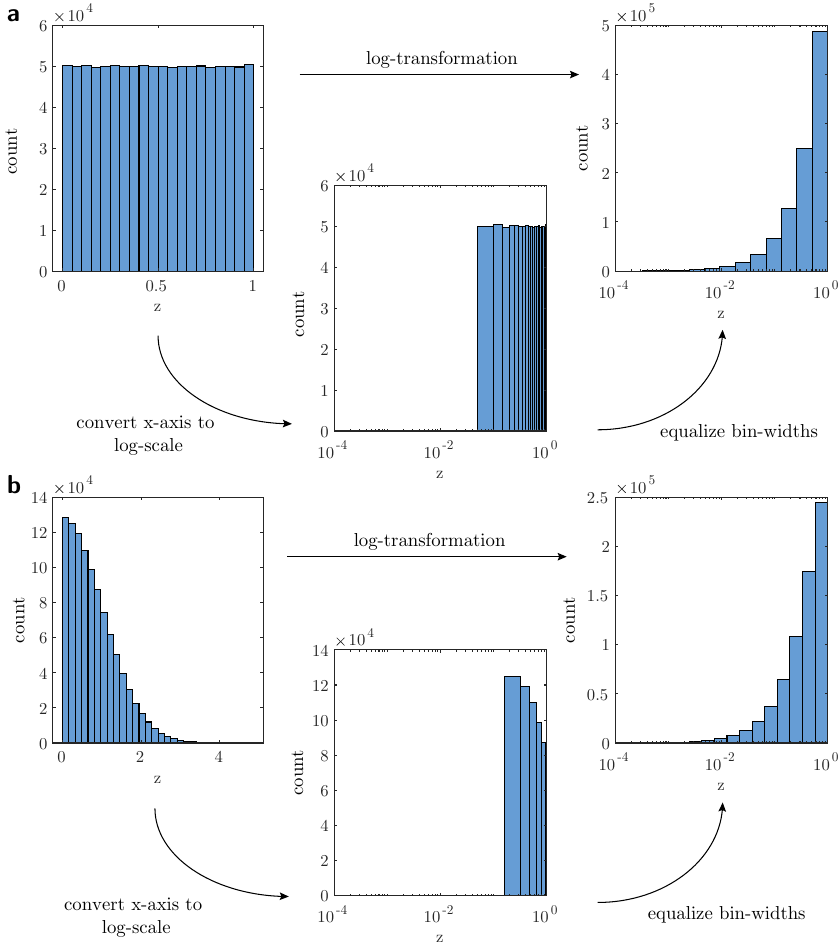}
\end{center}
\captionof{figure}{{\bf Procedure of $\boldsymbol{\log}$-transform.} The $\log$-transformation of the {\bf a,} uniform distribution and {\bf b,} half-normal distribution is equivalent to converting the x-axis to $\log$-scale followed by equalizing the bin-widths in the $\log$-space.}
\label{fig:extndfig1}

\begin{center}
  \includegraphics[width=\textwidth]{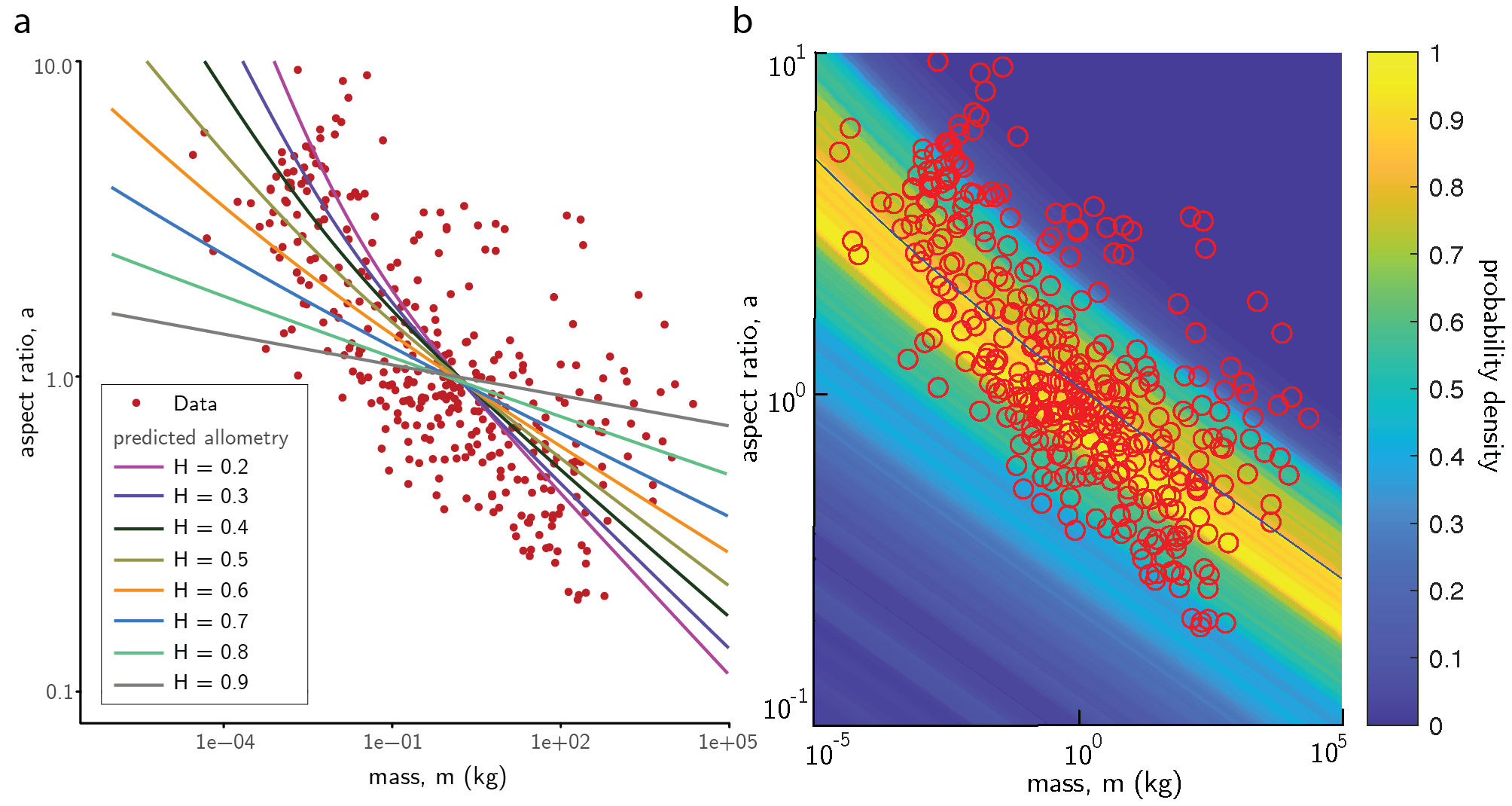}
\end{center}
\captionof{figure}{{\bfseries Scaling law for the varying value of Hurst exponent $H$ and Monte Carlo simulation of the probability densities.} {\bf a,} Predicted allometry for varying values of the Hurst exponent $H$ conserves the trend in the scaling of aspect ratio with mass. {\bf b,} Probability densities of aspect ratio and mass in the $\log$-space computed by propagating the distribution of the terrain height increments through the governing equations using Monte Carlo simulations.}
\label{fig:extndfig3}

\begin{center}
  \includegraphics[width=\textwidth]{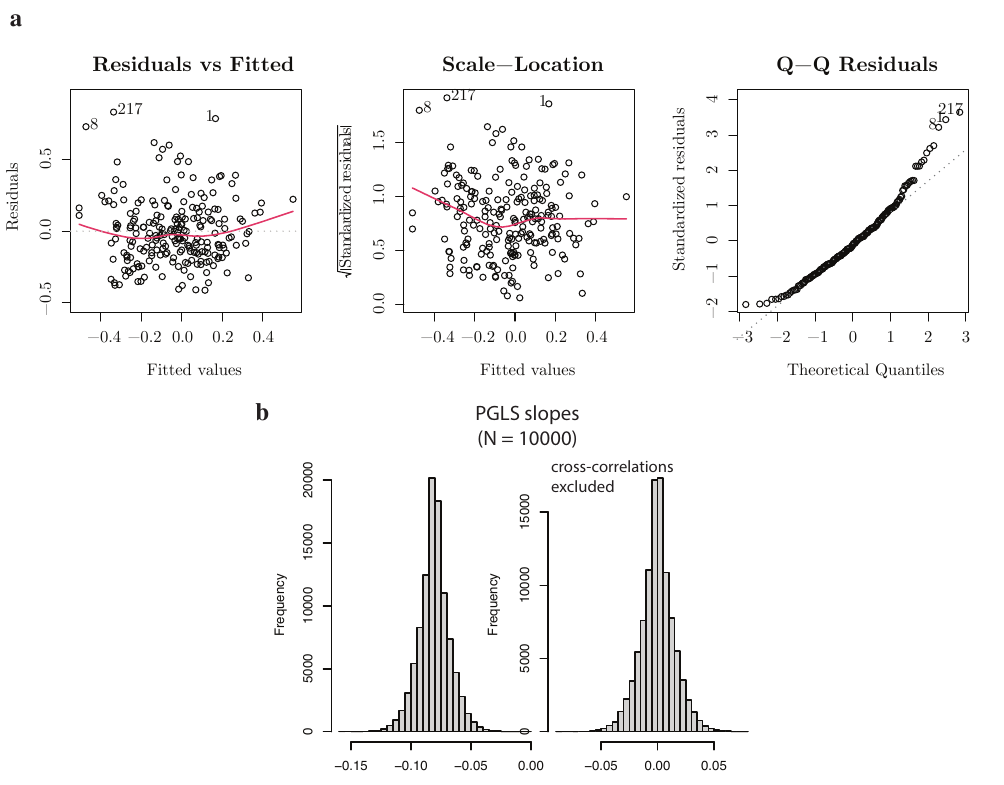}
\end{center}
\captionof{figure}{{\bfseries Analysis of data using linear regression and phylogenetic comparative methods.} {\bf a,} Diagnostics on the residuals obtained from ordinary linear regression show that the assumptions for linear regression are not met.
{\it Left,} residuals versus fitted plot does not suggest the presence of non linear patterns.
{\it Middle,} the scale location plot shows that the residuals violate the assumption of homoskedasticity (studentized Breusch-Pagan test for homoskedasticity: $p = 0.0792$, $BP = 3.0812$, $df = 1$).
{\it Right}, a Q-Q plot shows that the residuals are not normally distributed.
The dashed lines represent the ideal case when the linear regression assumptions would be met.
The red lines represent the local regression fits through the residuals.
{\bf b,} Histogram of slopes from simulated model of Brownian evolution on the phylogenetic tree under two different models of evolutionary trait correlations. The first model ({\it left,}\/) includes evolutionary cross-correlations whereas the second model ({\it right,}\/) excludes them.}
\label{fig:extndfig4}

\begin{center}
  \includegraphics[width=0.8\textwidth]{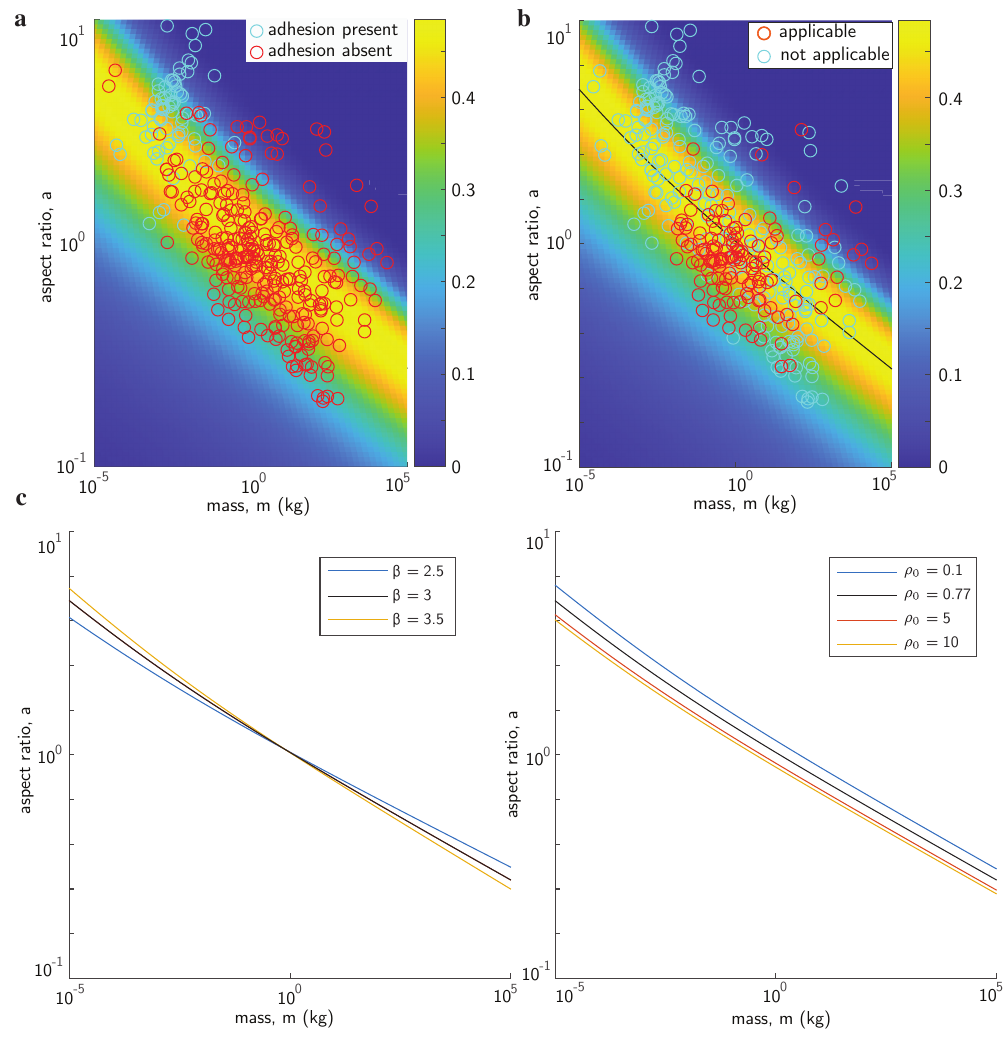}
\end{center}
\captionof{figure}{{\bfseries Analysis of biological variation.} {\bf a,} Animals sorted on the basis of adhesive pads in their limbs. {\bf b,} Animals sorted on based on the applicability of the parameters $\beta$ and $\rho_0$ calculated by Niklas, 1994 by modeling animals as cylinders. {\bf c,} Sensitivity analysis of the deterministic curve to the parameters $\beta$ and $\rho_0$ calculated by Niklas, 1994 shows that the predicted trend is preserved under a parameter range determined by biological variation.}
\label{fig:extndfig5}

\begin{center}
  \includegraphics[width=0.8\textwidth]{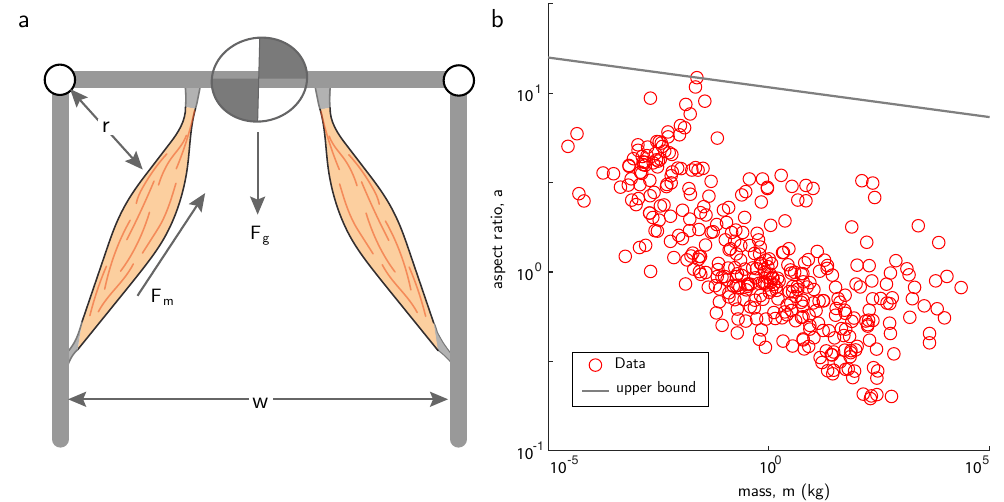}
\end{center}
\captionof{figure}{{\bfseries Upper bound based on muscle stress considerations.} {\bf a,} The frontal profile of an animal where the limbs are supported by muscles. As the width of the base increases, the moment arm of the ground reaction force about the center of mass increases and needs to be countered by muscle tension. {\bf b,} The upper bound scaling on the frontal aspect ratio obtained from the arguments based on peak skeletal and muscle stress considerations. The intercept is the fitting parameter, and we choose it so that all the data points are either lie on or below the upper bound.}
\label{fig:extndfig6}

\begin{center}
  \includegraphics[width=0.5\textwidth]{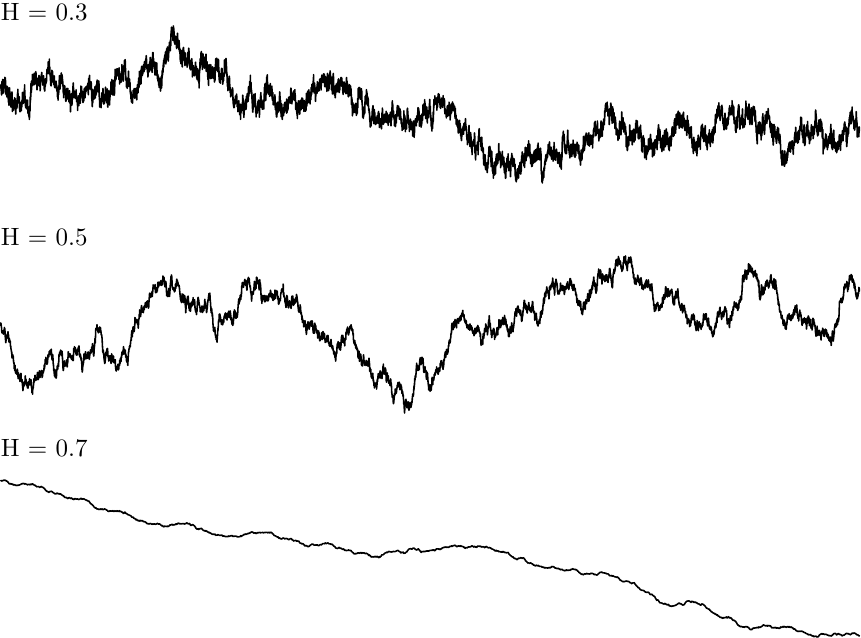}
\end{center}
\captionof{figure}{{\bfseries Simulation of terrain for varying values of $H$.} Simulation of fractional Brownian terrains with $H=0.3$, $H=0.5$, and $H=0.7$ shows the correlation between increments.}
\label{fig:extndfig7}

\clearpage
\section*{Extended Data Tables}

\captionof{table}{{\bf Information on data collected for 364 animals.} Selected entries from the complete table showing animal names, frontal aspect ratio, mass, and the source from where animal photograph and mass were obtained. Full table with 364 species available in ancillary file \texttt{extended\_data\_tables.pdf}.}\label{tab:extndtab1}
\begin{center}
\begin{tabular}{lcccc}
\toprule
\textbf{Animal} & \textbf{Aspect Ratio} & \textbf{Mass (kg)} & \textbf{Photo source} \\
\midrule
\textit{Chrysina macropus (m)} & 4.24 & 0.00286 & M Dickman \\
\textit{Elephas maximus} & 0.45 & 4360.00 & \cite{Wilson2011hoofed} \\
\textit{Bos bison} & 0.35 & 679.00 & \cite{Wilson2011hoofed} \\
\textit{Macrodontia cervicornis} & 7.66 & 0.01657 & M Dickman \\
\textit{Spheniscus mendiculus} & 0.65 & 2.60 & Photo Ark \\
\bottomrule
\multicolumn{4}{l}{\small Full dataset contains 364 species spanning 8 orders of magnitude in mass.}
\end{tabular}
\end{center}

\bigskip

\captionof{table}{{\bf Hurst exponent values from various terrain studies.} Summary of Hurst exponent $H$ measurements from the literature showing the range of values observed across different terrain types and measurement methods. Full table with over 200 terrain measurements available in ancillary file \texttt{extended\_data\_tables.pdf}.}\label{tab:extndtab2}
\begin{center}
\begin{tabular}{lccc}
\toprule
\textbf{Terrain type} & \textbf{H range} & \textbf{Scale} & \textbf{Reference} \\
\midrule
Natural landscapes & 0.3--0.8 & km & \cite{Goodchild:1982fk} \\
Soil surfaces & 0.4--0.7 & cm--m & \cite{Burrough1981} \\
Planetary surfaces & 0.5--0.9 & m--km & \cite{Shepard:2001kx} \\
Agricultural land & 0.3--0.6 & cm--m & \cite{Gallant1994} \\
\bottomrule
\end{tabular}
\end{center}

\end{document}